\documentclass{pasj00}

\begin{document}
\SetRunningHead{Hirota et al.}{Distance to Orion~KL}
\Received{}
\Accepted{}

\title{Distance to Orion~KL Measured with VERA}
\author{
Tomoya \textsc{Hirota},\altaffilmark{1,2}
Takeshi \textsc{Bushimata},\altaffilmark{1,3}
Yoon Kyung \textsc{Choi},\altaffilmark{1,4}
Mareki \textsc{Honma},\altaffilmark{1,2} \\
Hiroshi \textsc{Imai},\altaffilmark{5} 
Kenzaburo \textsc{Iwadate},\altaffilmark{6}
Takaaki \textsc{Jike},\altaffilmark{6}
Seiji \textsc{Kameno},\altaffilmark{5} \\
Osamu \textsc{Kameya},\altaffilmark{2,6}
Ryuichi \textsc{Kamohara},\altaffilmark{1} 
Yukitoshi \textsc{Kan-ya},\altaffilmark{7}
Noriyuki \textsc{Kawaguchi},\altaffilmark{1,2,3}\\
Masachika \textsc{Kijima},\altaffilmark{2}
Mi Kyoung \textsc{Kim},\altaffilmark{1,4}
Hideyuki \textsc{Kobayashi},\altaffilmark{1,3,4,6}
Seisuke \textsc{Kuji},\altaffilmark{6} \\
Tomoharu \textsc{Kurayama},\altaffilmark{1}
Seiji \textsc{Manabe},\altaffilmark{2,6}
Kenta \textsc{Maruyama},\altaffilmark{8}
Makoto \textsc{Matsui},\altaffilmark{8}\\
Naoko \textsc{Matsumoto},\altaffilmark{8}
Takeshi \textsc{Miyaji},\altaffilmark{1,3}
Takumi \textsc{Nagayama},\altaffilmark{8}
Akiharu \textsc{Nakagawa},\altaffilmark{5}\\
Kayoko \textsc{Nakamura},\altaffilmark{8}
Chung Sik \textsc{Oh},\altaffilmark{1,4}
Toshihiro \textsc{Omodaka},\altaffilmark{5}
Tomoaki \textsc{Oyama},\altaffilmark{1} \\
Satoshi \textsc{Sakai},\altaffilmark{6}
Tetsuo \textsc{Sasao},\altaffilmark{9,10}
Katsuhisa \textsc{Sato},\altaffilmark{6}
Mayumi \textsc{Sato},\altaffilmark{1,4}\\
Katsunori M. \textsc{Shibata},\altaffilmark{1,2,3}
Motonobu \textsc{Shintani},\altaffilmark{8}
Yoshiaki \textsc{Tamura},\altaffilmark{2,6}
Miyuki \textsc{Tsushima},\altaffilmark{8}\\
and Kazuyoshi \textsc{Yamashita}\altaffilmark{2}
}

\altaffiltext{1}{Mizusawa VERA Observatory, National Astronomical Observatory of Japan, \\
  2-21-1 Osawa, Mitaka, Tokyo 181-8588}
\altaffiltext{2}{Department of Astronomical Sciences, Graduate University for Advanced Studies, \\
  2-21-1 Osawa, Mitaka, Tokyo 181-8588}
\altaffiltext{3}{Space VLBI Project, National Astronomical Observatory of Japan, \\
  2-21-1 Osawa, Mitaka, Tokyo 181-8588}
\altaffiltext{4}{Department of Astronomy, Graduate School of Science, The University of Tokyo, \\
  7-3-1 Hongo, Bunkyo-ku, Tokyo 113-0033}
\altaffiltext{5}{Faculty of Science, Kagoshima University, \\
  1-21-35 Korimoto, Kagoshima, Kagoshima 890-0065}
\altaffiltext{6}{Mizusawa VERA Observatory, National Astronomical Observatory of Japan, \\
  2-12 Hoshi-ga-oka, Mizusawa-ku, Oshu-shi, Iwate 023-0861}
\altaffiltext{7}{Department of Astronomy, Yonsei University, \\
  134 Shinchong-dong, Seodaemun-gu, Seoul 120-749, Republic of Korea}
\altaffiltext{8}{Graduate School of Science and Engineering, Kagoshima University, \\
  1-21-35 Korimoto, Kagoshima, Kagoshima 890-0065}
\altaffiltext{9}{Department of Space Survey and Information Technology, Ajou University, \\
  Suwon 443-749, Republic of Korea}
\altaffiltext{10}{Korean VLBI Network, Korea Astronomy and Space Science Institute, \\
  P.O.Box 88, Yonsei University, 134 Shinchon-dong, Seodaemun-gu, Seoul 120-749, Republic of Korea}
\email{tomoya.hirota@nao.ac.jp}

\maketitle

\begin{abstract}
We present the initial results of multi-epoch VLBI observations of 
the 22~GHz H$_{2}$O masers in the Orion~KL region with VERA (VLBI Exploration of 
Radio Astrometry). With the VERA dual-beam receiving system, we have carried out 
phase-referencing VLBI astrometry and successfully detected an annual parallax of 
Orion~KL to be 2.29$\pm$0.10~mas, corresponding to the distance of 437$\pm$19 pc 
from the Sun. The distance to Orion~KL is determined for the first time with 
the annual parallax method in these observations. 
Although this value is consistent with that of the previously reported, 
480$\pm$80 pc, which is estimated from the statistical parallax method using 
proper motions and radial velocities of the H$_{2}$O maser features, 
our new results provide the much more accurate value with an uncertainty of only 4\%. 
In addition to the annual parallax, we have detected an absolute proper motion 
of the maser feature, suggesting an outflow motion powered by the radio source~I 
along with the systematic motion of source~I itself. 
\end{abstract}

\section{Introduction}

Distance is one of the most fundamental parameters in astronomy. 
However, it has been difficult to measure accurate distances to stars, galaxies, 
and other astronomical objects without assumptions. 
The most reliable way to determine the distance is an annual trigonometric parallax method, 
based on precise measurements of position and motion of the object. 
In 1990's, the Hipparcos satellite extensively measured annual parallaxes 
for more than 100~000 stars with a typical precision of 1~mas level (\cite{perryman1995}, 1997), 
which allowed us to refine various fields of astronomy and astrophysics. 
Nevertheless, the distances measured with Hipparcos were limited only within a few hundred pc 
from the Sun, which was far smaller than the size of the Galaxy, 15~kpc in radius. 

In the last decade, phase-referencing VLBI astrometry has been developed, 
with which the position of a target source is measured with 
respect to a reference source (\cite{beasley1995}). 
Using extragalactic radio sources as the position references 
(e.g. sources listed in the ICRF catalog; \cite{ma1998}), we can measure 
the absolute position of the target source, which lead us to derive its annual parallax. 
With recent highly precise VLBI astrometry, annual parallaxes have been 
successfully measured for the Galactic CH$_{3}$OH maser sources at the 12~GHz band 
(\cite{xu2006}) and H$_{2}$O maser sources at the 22~GHz band 
(\cite{kurayama2005}; \cite{hachisuka2006}) with the NRAO Very Long Baseline Array (VLBA). 
The annual parallax measurements with VLBI have also been carried out for non-thermal 
radio continuum emission from young stellar objects (e.g. \cite{lestrade1999}; \cite{loinard2005}). 
The highest accuracy of these VLBI astrometry is reported to be 0.05~mas, 
which provides a powerful tool to measure annual parallaxes with the accuracy by two orders 
of magnitude higher than that of the Hipparcos satellite, 
allowing us to measure the distances of maser sources up to 2~kpc away from the Sun 
(\cite{kurayama2005}; \cite{xu2006}; \cite{hachisuka2006}). 

In order to extend the VLBI astrometry of maser sources to the whole region 
of the Galaxy, we have constructed a new VLBI network in Japan called VERA, 
VLBI Exploration of Radio Astrometry (\cite{kobayashi2003}), 
which is the first VLBI array dedicated to phase-referencing observations. 
Each VERA antenna is equipped with a unique dual beam receiving system 
(\cite{kawaguchi2000}; \cite{honma2003}), which enables us to observe the target 
and reference sources within 2.2 degrees separation on the sky simultaneously, 
thus facilitating more efficient phase-referencing VLBI observations compared with 
the conventional fast-switching observations. 
Very recently, the first results of astrometry with VERA 
have been reported (e.g. \cite{honma2007}; \cite{sato2007}), 
demonstrating its high capability of annual parallax and absolute proper motion measurements. 
The main goal of the VERA project is to reveal 3-dimensional 
Galactic structure and kinematics based on the accurate astrometry 
of hundreds of H$_{2}$O (at the 22~GHz band) and SiO (at the 43~GHz band) 
maser sources in the Galactic star-forming regions and late-type stars 
with the highest accuracy of 10~$\mu$as level (\cite{kobayashi2003}; \cite{honma2000}). 

In this paper, we present the initial results of the annual parallax measurements of Orion~KL. 
Because Orion~KL is the nearest high-mass star-forming region 
located at an estimated distance of only 480 pc from the Sun (\cite{genzel1981}), 
it has been recognized as one of the most important objects to study high-mass 
star-formation processes (e.g. \cite{genzel1989}). Along with its proximity to the Sun, 
Orion~KL is known to be one of the brightest H$_{2}$O maser sources in the Galaxy, 
and hence, it is the best test bench for the first stage of the annual parallax measurements with VERA. 

\section{Observations and Data Analyses}

Observations of H$_{2}$O masers ($6_{1 6}$-$5_{2 3}$, 22235.080 MHz) in Orion~KL were  
carried out in 19 observing sessions from Jan. 2004 to Jul. 2006 with VERA. 
In this paper, we employed the results of total 16 observing sessions 
which were carried out under relatively good weather conditions. 
A typical interval of observations was 1 month, while some of them, especially 
in the summer season, were a few months. 
All the 4 stations of VERA were used in most of the observing sessions, 
while only 3 stations were used in part of the sessions 
(2004/027, 2004/272, and 2004/333; hereafter an observing session is denoted 
by year/day of the year). 
The maximum baseline length was 2270~km (see Fig.1 of \cite{petrov2007}) 
and the typical synthesized beam size 
(FWMH) was 1.5~mas$\times$0.8~mas with a position angle of $-30$~degrees. 

All the observations were made in the dual beam mode; Orion~KL and an ICRF source 
J0541$-$0541 ($\alpha(J2000)=$05h41m38.083385s, $\delta(J2000)=-05$d41'49.42839"; 
\cite{ma1998}; \cite{petrov2007}) were observed simultaneously. 
The separation angle between them was 1.62~degrees. 
J0541$-$0541 was detected fringes with a flux density of about 500~mJy in all the 
observations, which was suitable as a phase reference source. 
The instrumental phase difference between the two beams was measured 
in real time during the observations, using the correlated data of the random signal 
from artificial noise sources injected into two beams at each station (\cite{kawaguchi2000}). 
The typical value of the phase drift between the two beams was 3 degrees per hour. 
These results were used for calibrating instrumental effects in the observed 
phase difference between the two sources. 

Left-handed circular polarization was received and sampled with 2-bit 
quantization, and filtered using the VERA digital filter unit (\cite{iguchi2005}). 
The data were recorded onto magnetic tapes at a rate of 1024~Mbps, 
providing a total bandwidth of 256~MHz in which one IF channel and the rest 
of 15 IF channels with 16~MHz bandwidth each were assigned to 
Orion~KL and J0541$-$0541, respectively. 
In the earlier eight observing sessions from 2004/203 to 2005/144, we used 
the recording system at a rate of 128~Mbps, with two IF channels of 
16~MHz bandwidth each for both Orion~KL and J0541$-$0541. A bright continuum source, 
J0530+1331, was observed every 1-2 hours for bandpass and delay calibration. 
System temperatures including atmospheric attenuation were measured with 
the chopper-wheel method (\cite{ulich1976}) to be 100-600 K, 
depending on weather conditions and elevation angle of the observed sources. 
The aperture efficiencies of the antennas ranged from 45 to 52\% depending on 
the stations. 
A variation of the aperture efficiency of each antenna as a function of elevation angle 
was confirmed to be less than 10\% even at the lowest elevation in the observations 
($\sim$20~degrees). 

Correlation processing was carried out on the Mitaka FX correlator 
(\cite{chikada1991}) located at the NAOJ Mitaka campus. 
For H$_{2}$O maser lines, a spectral resolution was set to be 15.625~kHz, 
corresponding to the velocity resolution of 0.21~km~s$^{-1}$. 
The effective velocity coverage for the H$_{2}$O maser lines, 
which was common for all the observing sessions, 
was $\pm40$~km~s$^{-1}$ relative to the systemic velocity of Orion~KL, 
an LSR velocity of 8~km~s$^{-1}$. 

Calibration and imaging were performed using 
the NRAO Astronomical Image Processing System (AIPS). 
At first, amplitude and bandpass calibration were done for each target 
(Orion~KL) and reference source (J0541$-$0541) independently. 
Then fringe fitting was made with the AIPS task FRING on 
the phase reference source (J0541$-$0541), and 
the phase solutions were applied to the target source (Orion~KL). 
In addition, we adopted the results of dual-beam phase calibration measurements 
as described above (\cite{kawaguchi2000}). 
Because the a priori delay model applied in the correlation processing 
was not accurate enough for precise astrometry, 
we calibrated the visibility phase using the more accurate delay model, 
based on the recent achievements of geodynamics (\cite{honma2007}) in the analyses. 
In this model, we calibrated the fluctuation of the visibility phase 
caused by the Earth's atmosphere based on the GPS measurements of 
the atmospheric zenith delay due to the tropospheric water vapor. 

The synthesized images were made using the AIPS task IMAGR with natural weighting. 
Even after the phase calibrations described above, we found that the dynamic range of 
the phase-referenced images was not high enough, possibly due to a residual in 
the atmospheric zenith delay, as pointed out by \citet{honma2007}. 
To improve the quality of these images, we estimated 
the atmospheric zenith delay residual as a constant offset for 
each station, which maximized the coherence of the resultant phase-referenced image. 
The atmospheric zenith delay residual was derived to be 0-10~cm on average, depending 
on the weather conditions, while it exceeded 20~cm in the worst case. 
As a result of this calibration, the dynamic range of each phase-referenced image 
was increased by a factor of up to 1.5. 

\section{Results}

\begin{figure}
  \begin{center}
     \FigureFile(80mm,80mm){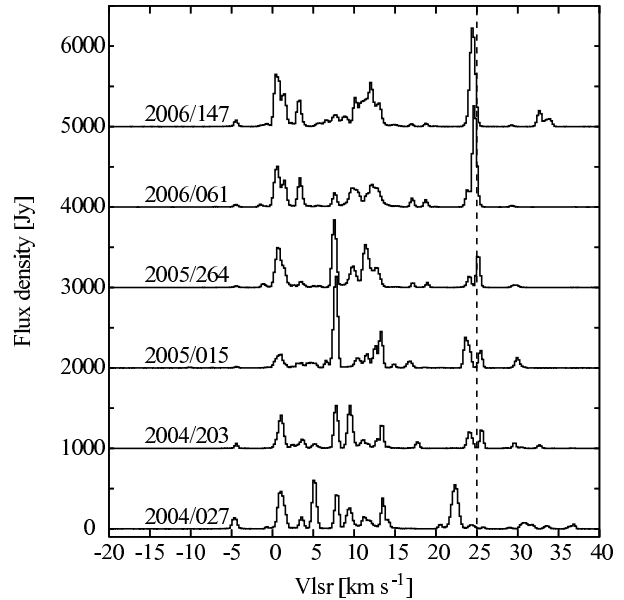}
  \end{center}
  \caption{Examples of scalar-averaged cross power spectra of Orion~KL observed 
   with the VERA Mizusawa-Iriki baseline (1267~km). A dashed line indicates the central velocity 
   of the maser feature adopted for the parallax measurement in this paper 
   at the LSR velocity of 25~km~s$^{-1}$. }
  \label{fig-spectra}
\end{figure}

Figure \ref{fig-spectra} shows the cross power spectra of the 
H$_{2}$O masers toward Orion~KL. The H$_{2}$O maser lines were detected 
within the LSR velocity range from $-10$~km~s$^{-1}$ to 40~km~s$^{-1}$. 
We could not find high-velocity components 
in the LSR velocity of $>40$~km~s$^{-1}$ and $<-10$~km~s$^{-1}$ 
(\cite{genzel1981}) possibly due to our narrower effective 
velocity coverage (from $-32$~to~48~km~s$^{-1}$) 
and lower sensitivity. 

\begin{figure}
  \begin{center}
    \FigureFile(80mm,80mm){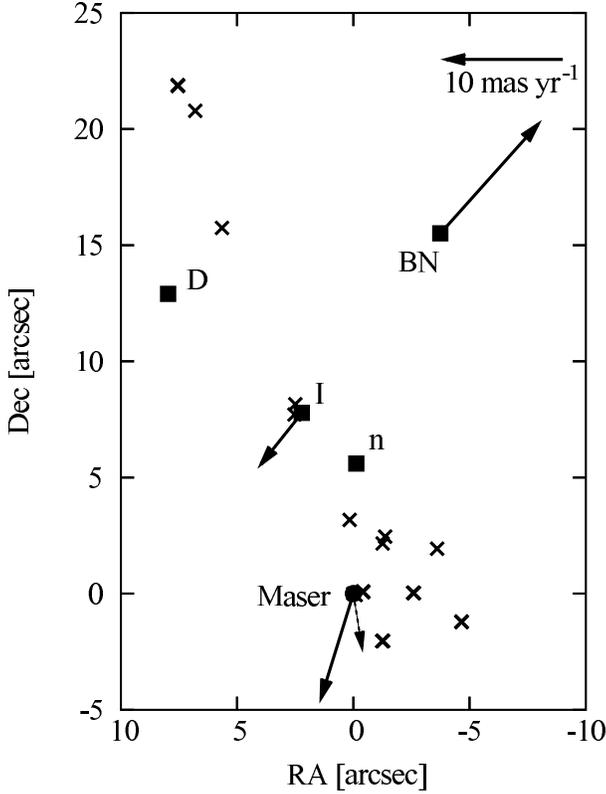}
  \caption{Distribution of H$_{2}$O masers in Orion~KL observed in the epoch of 2005/081. 
   Crosses represent the positions of individual or groups of H$_{2}$O maser features. 
   Filled squares and circle indicate the positions of radio continuum sources (\cite{gomez2005}) 
   and the maser feature analyzed in this paper at the LSR velocity of 24.5-25.7~km~s$^{-1}$. 
   Bold arrows indicate the absolute proper motion vectors based on our study and \citet{rodriguez2005}, 
   while a dashed arrow shows the proper motion of the maser feature 
   with respect to source~I (see text). 
   The position offsets are with respect to the reference position 
   ($\alpha(J2000)=$05h35m14.363600s, $\delta(J2000)=-05$d22'38.30100"). }
  \label{fig-map}
  \end{center}
\end{figure}

In order to reveal the overall distribution of the H$_{2}$O masers, 
we at first mapped the H$_{2}$O maser features in the Orion~KL region 
at one of the observed sessions, 2005/081, by the method adopted in 
usual single-beam VLBI observations. 
The H$_{2}$O maser features are found to be extended over 
the 20"$\times$30" region as shown in Figure \ref{fig-map}. 
The distribution of H$_{2}$O maser features is in good agreement with those 
in \citet{genzel1981} and \citet{gaume1998}. 
The number of H$_{2}$O maser features near source~I, which is proposed 
to be a powering source of the outflow and the H$_{2}$O masers 
(\cite{menten1995}; \cite{greenhill1998}), 
is smaller than that of the results of the NRAO Very Large Array 
(VLA) observations reported by \citet{gaume1998}. 
This is because most of the maser features near source~I are 
resolved out with the synthesized beam of VERA, implying that their sizes are 
larger than a few mas (\cite{genzel1981}; \cite{gaume1998}). 

Based on the H$_{2}$O maser map at the epoch of 2005/081, 
we searched for intense H$_{2}$O maser features whose cross power 
spectra observed with the Mizusawa-Iriki baseline (1267~km; see Fig.1 of \cite{petrov2007}) 
were detected with a signal to noise ratio larger than 10 at all the 16 observing epochs. 
We found that 10 maser features satisfied this criterion. 
Among them, we analyzed the data for one of the maser features 
at the LSR velocity of about 25~km~s$^{-1}$, which was redshifted relative to 
that of the systemic velocity of Orion~KL, an LSR velocity of 8~km~s$^{-1}$, 
showing relatively less significant spatial structure in the synthesized 
images and the closure phases during all the observing sessions. 
Since the peak velocity of the maser feature was shifted systematically 
from 25.7~km~s$^{-1}$ to 24.5~km~s$^{-1}$ during the observing period of 2~years, 
we made images of maser spots for all the spectral channels within 
the velocity range of 24.5-25.7~km~s$^{-1}$, and determined the position of 
the maser feature taking that of the peak velocity channel. 
Although we cannot rule out the possibility of acceleration of this maser feature, 
the observed velocity shift would imply a variation of its source structure. 
Along with the velocity shift, 
the flux density of the maser feature was also highly variable as shown in Figure \ref{fig-spectra}. 
The variation of the maser feature suggested by the velocity shift and the flux variability 
would affect the accuracy of astrometry even if the maser is 
bright, relatively stable, and with less significant spatial structure, as described later. 
Detailed analyses for all the H$_{2}$O maser features will be reported in a forthcoming paper. 

\begin{figure*}
  \begin{center}
    \FigureFile(160mm,160mm){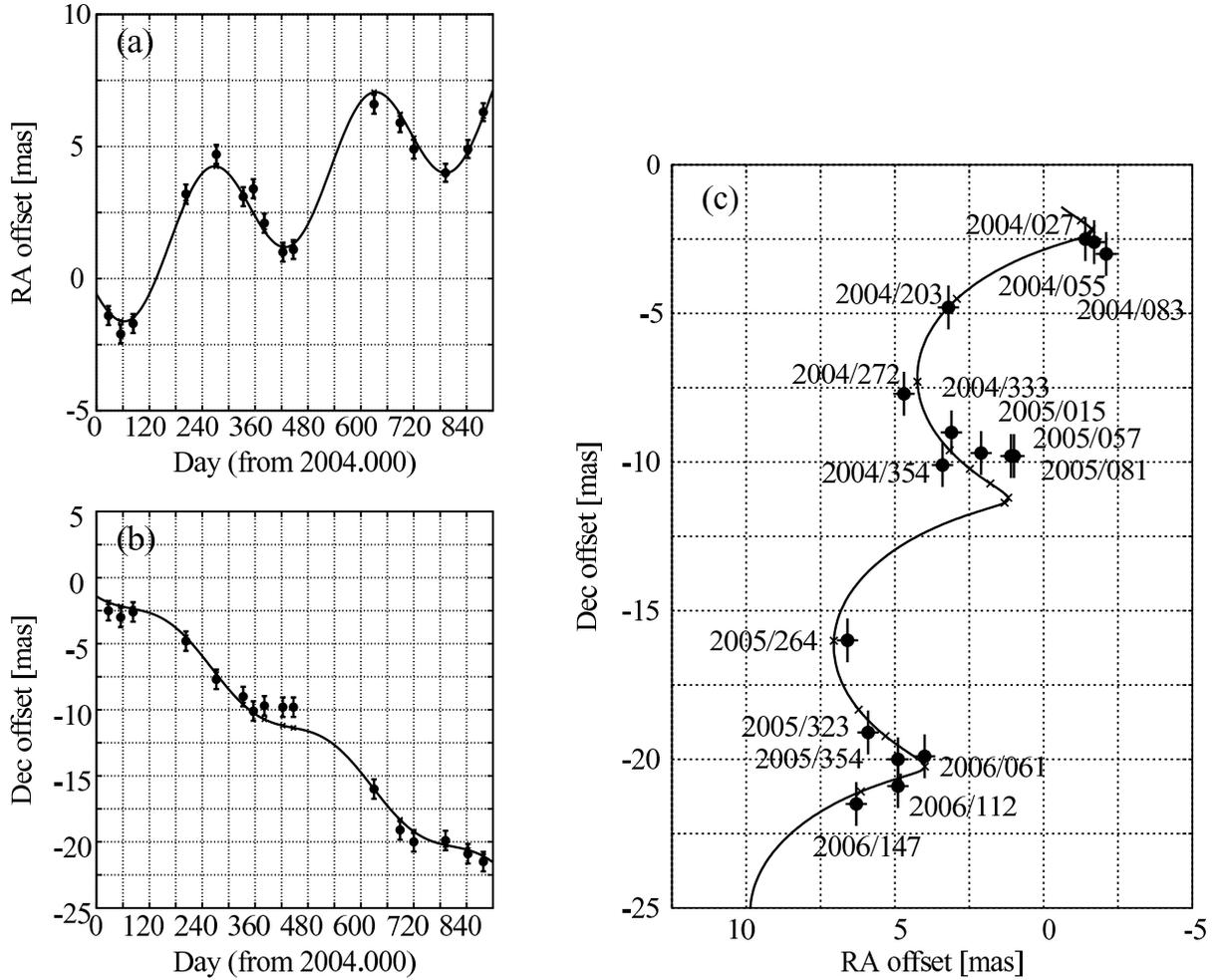}
  \caption{Results of the position measurements of the maser feature in Orion~KL. 
   (a) The movement of the maser feature in right ascension as a function of time. 
   (b) The same as (a) in declination. 
   (c) The movement of the maser feature on the sky. Solid lines represent the 
   best fit model with the annual parallax and linear proper motion for the 
   maser feature. Filled circles represent the observed positions of the maser 
   feature with error bars indicating the standard deviations of the 
   least-squares analysis as listed in Table \ref{tab-results} (0.36~mas in right ascension 
   and 0.74~mas in declination). 
   The reference position is the same as in Figure \ref{fig-map}. 
   Observed epochs are indicated in the panel (c). }
  \label{fig-ra}
  \end{center}
\end{figure*}

Figure \ref{fig-ra} gives the results of our position measurements of 
the H$_{2}$O maser feature. 
As shown in Figure \ref{fig-ra}, we have successfully measured the movement of the H$_{2}$O maser 
feature for longer than 2~years. 
The movement significantly deviates from a linear motion showing a sinusoidal 
modulation with a period of 1~year. This is clearly due to the 
annual parallax of the maser feature. In fact, the date of each peak in the sinusoidal curve 
is almost consistent with those predicted from the annual parallax of Orion~KL. 
Assuming that the movement of the maser feature is the sum of 
linear motion and the annual parallax, we can obtain the proper motion in right ascension 
$\mu_{\alpha}$ and declination $\mu_{\delta}$, the initial position in right ascension $\alpha_{0}$
and declination $\delta_{0}$, and the annual parallax $\pi$ for the maser feature by 
a least-squares analysis. 

Initially, we determined these 5 parameters simultaneously, 
using both right ascension and declination data. In this case, 
the derived annual parallax was 2.25$\pm$0.21~mas, corresponding to 
the distance of 445$\pm$42~pc, and the standard deviations of the 
least-squares analysis in right ascension $\sigma_{\alpha}$ and in declination $\sigma_{\delta}$ 
were 0.36~mas and 0.74~mas, respectively. The larger standard deviation in declination 
suggests that the astrometric accuracy in the declination is significantly worse 
than that in the right ascension. This trend can be seen in other observations 
with VERA (\cite{honma2007}; \cite{sato2007}). 
One of the possible reasons for this is that the residual of the atmospheric zenith delay would 
affect the astrometric accuracy, as discussed later. 
Therefore, we at first determined the absolute proper motion and initial position 
in right ascension together with the annual parallax using the data for 
right ascension only. 
As a result, we obtained the annual parallax with higher precision to be 
2.29$\pm$0.10~mas, corresponding to the distance of 437$\pm$19~pc. 
After the annual parallax was derived from the right ascension data, 
we estimated the absolute proper motion and initial position in declination 
using the data for declination. 
The results are summarized in Table \ref{tab-results}. 

\begin{table}[tb]
\begin{center}
\caption{Results of the least-squares analysis for the annual parallax and proper motion measurements}
\label{tab-results}
\begin{tabular}{lc}
 \hline\hline
Parameter                         & Best fit value               \\ 
\hline
$\pi$                             &  2.29(0.10)~mas              \\
$\mu_{\alpha}$                    &  2.77(0.09)~mas~yr$^{-1}$    \\
$\mu_{\delta}$                    & $-8.97(0.21)$~mas~yr$^{-1}$  \\
$\sigma_{\alpha}$                 &  0.36~mas                    \\
$\sigma_{\delta}$                 &  0.74~mas                    \\
  \hline \\
\multicolumn{2}{l}{Note --- Numbers in parenthesis represent the } \\
\multicolumn{2}{l}{\quad estimated uncertainties. Annual parallax $\pi$ is } \\
\multicolumn{2}{l}{\quad derived from the right ascension data only. } \\
\end{tabular}
\end{center}
\end{table}

\section{Discussions}

\begin{table*}[thb]
\begin{center}
\caption{Results of the proper motion measurements for the observed maser feature and source~I}
\label{tab-proper}
\begin{tabular}{lccccccccc}
 \hline\hline
 & \multicolumn{4}{c}{Absolute proper motion} & & \multicolumn{4}{c}{Proper motion relative to source~I} \\
 \cline{2-5}  \cline{7-10} & & & \\
Source        & $\mu_{\alpha}$ & $\mu_{\delta}$ & $\mu$ & $v_{t}$ & &
 $\mu_{\alpha}^{I}$ & $\mu_{\delta}^{I}$ & $\mu^{I}$ & $v_{t}^{I}$   \\
Name          & (mas~yr$^{-1}$) & (mas~yr$^{-1}$) & (mas~yr$^{-1}$) & (km~s$^{-1}$) & &
(mas~yr$^{-1}$) & (mas~yr$^{-1}$) & (mas~yr$^{-1}$) & (km~s$^{-1}$)  \\
\hline
Maser$^{a}$ & 2.77(0.09)   & $-8.97(0.21)$ & 9.39(0.20) & 19.7(0.4)$^{b}$ & &
               $-0.7(0.7)$ & $-4.6(0.7)$ & 4.6(0.7) &  9.7(1.5)$^{b}$  \\
source~I$^{a}$ & 3.5(0.7)   & $-4.4(0.7)$   & 5.6(0.7)    & 12(2)$^{b}$     & &
               0.00         & 0.00          & 0.00       & 0.00        \\
\hline \\
\multicolumn{10}{l}{Note --- Numbers in parenthesis represent the estimated uncertainties. } \\
\multicolumn{10}{l}{$a$: Absolute proper motion of source~I is taken from \citet{rodriguez2005}. }\\
\multicolumn{10}{l}{$b$: Calculated assuming the distance of 437~pc. }
\end{tabular}
\end{center}
\end{table*}

\subsection{Astrometric error sources}

In this paper, we successfully measured the annual parallax of Orion~KL 
to be 2.29$\pm$0.10~mas through the 2-year monitoring observations of the H$_{2}$O 
maser feature with VERA. 
The sinusoidal curve of the movement of the maser feature 
as shown in Figure \ref{fig-ra} is almost coincident with the predicted annual parallax 
of Orion~KL both in period (1~year) and phase (date of the peaks in the sinusoidal curve). 
Therefore, the deviation from the best fit model, which is the combination of 
annual parallax and linear proper motion of the maser feature, should be regarded 
as astrometric errors in our observations, rather than due to an inappropriate model 
in the least-squares analysis. 
In this section, we will consider possible sources of these astrometric errors. 

As reported previously in the literature 
(\cite{kurayama2005}; \cite{hachisuka2006}; \cite{honma2007}; \cite{sato2007}), 
it is difficult to estimate the individual error sources in the VLBI astrometry quantitatively. 
We therefore estimate the uncertainties in the measured position of the maser feature 
to be 0.36~mas and 0.74~mas in right ascension and declination, respectively, 
based on the standard deviations of the least-squares analysis as listed in 
Table \ref{tab-results}. The standard deviations 
obtained in this paper are larger than those of previous observations with VERA 
(\cite{honma2007}; \cite{sato2007}), especially in declination. 

The most serious error source in the VLBI astrometry in the 22~GHz band 
is likely to be the atmospheric zenith delay residual due to the tropospheric 
water vapor. 
This is caused by the difference in the optical path lengths through 
the atmosphere between the target and reference sources 
because the elevation angle of the target source is usually different from 
that of the reference source. 
According to the discussions in \citet{honma2007}, a path length error due to 
the atmospheric zenith delay residual of 3~cm would cause a position error 
of 0.04-0.12~mas in the case of a separation angle between the target and 
reference sources of 0.7~degrees at the elevation angle of 20-90~degrees. 
If we consider an extreme example, with the observed elevation angle of 
20~degrees and the atmospheric zenith delay residual of 10~cm, 
the position error in the observations of Orion~KL and J0541$-$0541, with 
a separation angle of 1.62~degrees, is estimated to be 0.75~mas. 
This value is clearly overestimated because the path length errors should be suppressed 
at the higher elevation angle. Furthermore, the atmospheric zenith delay residual of 
10~cm is unrealistic because we have corrected such a large residual before 
phase-referencing imaging. 
Therefore, the atmospheric zenith delay residual alone cannot fully explain our 
position errors, although it would contribute to the large part of the error 
source in our astrometry, especially in declination. 

One of other possibilities for the error sources in the observed position 
is a variation of the structure in the maser feature. With regard to this, 
we confirmed that peak positions of the maser spots within the analyzed maser feature 
were sometimes shifted by about 0.2~mas from those of the adjacent channels. 
In addition, the systematic velocity shift from 25.7~km~s$^{-1}$ to 24.5~km~s$^{-1}$ 
was observed during the observing period of 2~years, 
indicating the variation of the maser feature. 
Although there is no reason that the structure in the maser feature affects 
the astrometric accuracy only in declination, it would be one of the major 
sources of errors in the astrometry with the H$_{2}$O maser lines as well as 
the atmospheric zenith delay residual. 
The effect of the spatial structure of the maser feature is more significant for Orion~KL than 
the other sources (\cite{kurayama2005}; \cite{hachisuka2006}; \cite{honma2007}; \cite{sato2007}) 
because the distance to Orion~KL (437~pc) is nearer than the others by a factor of 5-10 (2-5~kpc). 
However, this effect is inversely proportional to the distance to the target source 
just the same as its annual parallax. 
This means that the annual parallaxes of the more distant sources can be measured 
with almost the same precision as in the case of Orion~KL, 
if the dominant error source in astrometry is due to the structure effect 
rather than the atmospheric zenith delay residual. 
In fact, the relative uncertainty in the annual parallax of the further source, S269, 
is found to be comparable to that of Orion~KL, about 4\%, in the case of using the data for 
right ascension only (\cite{honma2007}). 
Further VLBI observations of maser features with shorter baselines should be able to 
confirm this effect, with which more extended structures of maser features are imaged. 

On the other hand, the variation of the structure of the reference source, J0541$-$0541, 
would be negligible for the measurements of the annual parallax and proper motion because we 
found no evidence for significant structure of J0541$-$0541 in our observations. 
The uncertainty in the absolute position of the reference source J0541$-$0541, 
0.28~mas and 0.46~mas in right ascension and 
declination, respectively (\cite{ma1998}), also does not affect 
the derived annual parallax and proper motion because this uncertainty 
gives only a constant offset to the position of the maser feature. 
According to the discussions in \citet{honma2007}, astrometric errors in the VERA 
observations arising from uncertainties in the station position, delay model, 
and path length errors due to ionosphere are estimated to be smaller by an 
order of magnitude, and hence, they do not have significant effects on astrometric accuracy. 
Therefore, we conclude that the major sources of our astrometric errors are 
due to the atmospheric zenith delay residual and variability of the structure of the maser feature. 

\subsection{Annual Parallax and Distance to Orion~KL}

We successfully obtained the annual parallax 
of Orion~KL to be 2.29$\pm$0.10~mas, corresponding to the distance of 437$\pm$19 pc. 
This is the first time that the distance to Orion~KL is determined 
based on the annual parallax measurements. 
\citet{genzel1981} derived the distance to Orion~KL to be 480 $\pm$ 80 pc 
from the statistical parallax method, using proper motions and 
radial velocities of the H$_{2}$O maser features. 
Our result is consistent with that of \citet{genzel1981}, 
although the accuracy of our measurements is significantly improved. 
The most important progress in our new results is due to the geometric nature 
of our measurements without any assumption unlike the statistical parallax method, 
in which appropriate kinematic modeling for Orion~KL is required (\cite{genzel1981}). 
The accuracy of the annual parallax measurements in our study is limited mainly 
due to the atmospheric zenith delay residual and the structure of the maser feature, 
both of which are difficult to be predicted and measured completely in the current observational study. 
In principle, it will be possible to achieve much higher precision using the results 
of all the maser features in Orion~KL, which will reduce the statistical error by a factor of 
$N^{-0.5}$ where $N$ represents the number of observed maser features. 
This expectation will be confirmed in the further analyses of the VERA observations. 

\subsection{Absolute Proper Motion of the Maser feature in Orion~KL}

Along with the annual parallax measurements, we successfully detected 
the absolute proper motion in our phase-referencing astrometry with VERA. 
Figure \ref{fig-map}, Tables \ref{tab-results} and \ref{tab-proper} 
show the absolute proper motion of the maser feature in Orion~KL. 
At the distance of 437~pc, the proper motion of 1~mas~yr$^{-1}$ corresponds 
to the transverse velocity of 2.1~km~s$^{-1}$. 
The observed absolute proper motion of the H$_{2}$O maser feature 
(2.77$\pm$0.09~mas~yr$^{-1}$ and $-8.97\pm$0.21~mas~yr$^{-1}$ in right 
ascension and declination, respectively) corresponds to 9.39$\pm$0.20~mas~yr$^{-1}$ 
or 19.7$\pm$0.4~km~s$^{-1}$ toward south. 

Recently, \citet{rodriguez2005} and \citet{gomez2005} measured the proper motion 
of radio continuum sources in the Orion~KL region with the VLA, as shown in 
Figure \ref{fig-map} and Table \ref{tab-proper}. 
Subtracting the proper motion vector of source~I from that of the observed 
maser feature, we can obtain the proper motion of the maser feature with respect to source~I. 
As \citet{gomez2005} have already mentioned, the precision of the absolute 
proper motion measurements by \citet{rodriguez2005} is higher than that by \citet{gomez2005}. 
Therefore, we adopt the proper motion of source~I 
inferred by \citet{rodriguez2005}, 3.5$\pm$0.7~mas~yr$^{-1}$ and $-4.4\pm0.7$~mas~yr$^{-1}$ 
in right ascension and declination, respectively, in the following discussions. 
The proper motion of the maser feature with respect to source~I is inferred to be 
$-0.7\pm$0.7~mas~yr$^{-1}$ and $-4.6\pm0.7$~mas~yr$^{-1}$ 
in right ascension and declination, respectively, as listed in Table \ref{tab-proper}. 
The magnitude of the proper motion is 4.6$\pm$0.7~mas~yr$^{-1}$ or 
9.7$\pm$1.5~km~s$^{-1}$ toward south with a position angle of $-171$~degrees, 
which agrees well with the direction of the outflow powered by source~I. 
Therefore, we conclude that the absolute proper motion of the observed maser 
feature is the sum of outflow motion powered by source~I and the 
systematic motion of source~I itself. 

However, a detailed model of the outflow powered by source~I is still debatable. 
\citet{greenhill1998} first proposed that the biconical high-velocity outflow traced 
by the SiO maser lines lies along the northwest-southeast direction, 
while the low-velocity equatorial outflow traced by the H$_{2}$O maser lines 
exists along the northeast-southwest direction. 
On the other hand, they changed the interpretation based on the recent results 
that the outflow is along the northeast-southwest direction, which is perpendicular to 
the first model, and that the SiO maser lines trace 
the edge-on disk perpendicular to the outflow (\cite{greenhill2004}). 
We cannot distinguish these two different models in this paper because 
the distribution of the H$_{2}$O masers, elongated along the northeast-southwest 
direction as shown in Figure \ref{fig-map}, is consistent with both models 
and in addition, the proper motion of the observed H$_{2}$O maser feature is almost 
intermediate (toward south) between the proposed outflow axes (\cite{greenhill1998}, 2004). 
The velocity structure in the Orion~KL region is quite complicated as \citet{greenhill2004} 
suggested, and hence, further discussions about the proper motions 
of all the H$_{2}$O maser features are required to construct the detailed model of the outflow 
in the Orion~KL region, which will be presented in a forthcoming paper. 

\vspace{12pt}
The authors thank Dr. Yoshiaki Hagiwara for useful discussions and 
careful reading of the manuscript. 
We are also grateful to the anonymous referee for helpful comments and suggestions. 
TH is financially supported by Grant-in-Aids from 
the Ministry of Education, Culture, Sports, Science and Technology (13640242 and 16540224).


\begin{thebibliography}{}
\bibitem[Beasley \& Conway (1995)]{beasley1995}
  Beasley, A. J., Conway, J. E. \ 1995, in ASP Conf. Ser. 82, 
  Very Long Baseline Interferometry and the VLBA, ed. J. A. Zensus, P. J. Diamond, \& 
  P. J. Napier (San Francisco: ASP), 327
\bibitem[Chikada et al. (1991)]{chikada1991} Chikada, Y., et al. 
  1991, in Frontiers of VLBI, ed. H. Hirabayashi, M. Inoue, \& H. Kobayashi 
  (Tokyo: Universal Academy Press), 79 
\bibitem[Gaume et al. (1998)]{gaume1998}
  Gaume, R. A., Wilson, T. L., Vrba, F. J., Johnston, K. J., 
  Schmid-Burgk, J. \ 1998, ApJ, 493, 940
\bibitem[Genzel et al. (1981)]{genzel1981}
  Genzel, R., Reid, M. J., Moran, J. M., Downes, D. \ 1981, ApJ, 244, 884 
\bibitem[Genzel \& Stutzki (1989)]{genzel1989}
  Genzel, R., Stutzki, J. \ 1989, ARA\&A, 27, 41 
\bibitem[G\'omez et al. (2005)]{gomez2005}
   G\'omez, L., Rodr\'\i guez, L. F., Loinard, L., Lizano, S., 
   Poveda, A., Allen, C. \ 2005, ApJ, 635, 1166
\bibitem[Greenhill et al. (1998)]{greenhill1998}
   Greenhill, L. J., Gwinn, C. R., Schwartz, C., Moran, J. M., 
   Diamond, P. J. \ 1998, Nature, 396, 650
\bibitem[Greenhill et al. (2004)]{greenhill2004}
   Greenhill, L. J., Reid, M. J., Chandler, C. J., Diamond, P. J., 
   Elitzur, M. \ 2004, in IAU Symp. 221, Star Formation at High Angular Resolution, 
   ed. M. G. Burton, R. Jayawardhana, \& T. L. Bourke (San Francisco: ASP), 155
\bibitem[Hachisuka et al. (2006)]{hachisuka2006} 
  Hachisuka, K. et al. \ 2006, ApJ, 645, 337
\bibitem[Honma et al. (2000)]{honma2000} Honma, M., Kawaguchi, N., Sasao, T. \ 2000, 
  in Proc. SPIE, 4015, Radio Telescopes, ed. H. R. Butcher (Washington: SPIE), 624
\bibitem[Honma et al. (2003)]{honma2003} Honma, M. et al. \ 2003, PASJ, 55, L57
\bibitem[Honma et al. (2007)]{honma2007} Honma, M. et al. \ 2007, PASJ, in press
\bibitem[Iguchi et al. (2005)]{iguchi2005} Iguchi, S., Kurayama, T., 
  Kawaguchi, N., Kawakami, K. \ 2005, PASJ, 57, 259
\bibitem[Kawaguchi et al. (2000)]{kawaguchi2000} Kawaguchi, N., Sasao, T., Manabe, S. \ 2000, 
  in Proc. SPIE, 4015, Radio Telescopes, ed.\ H. R. Butcher (Washington: SPIE), 544
\bibitem[Kobayashi et al. (2003)]{kobayashi2003} 
  Kobayashi, H. et al. \ 2003, in ASP Conf. Ser. 306, 
  New technologies in VLBI, ed. Y.C. Minh, (San Francisco: ASP), 367
\bibitem[Kurayama et al. (2005)]{kurayama2005}
  Kurayama, T., Sasao, T., Kobayashi, H. \ 2005, ApJ, 627, L49
\bibitem[Loinard et al. (2005)]{loinard2005}
  Loinard, L., Mioduszewski, A. J., Rodr\'\i guez, L. F., Gonz\'alez, R. A., 
  Rodr\'\i guez, M. I., Torres, R. M. \ 2005, ApJ, 619, L179
\bibitem[Lestrade et al. (1999)]{lestrade1999}
  Lestrade, J.-F., Preston, R. A., Jones, D. L., Phillips, R. B., Rogers, A. E. E., 
  Titus, M. A., Rioja, M. J., Gabuzda, D. C. \ 1999, A\&A, 344, 1014
\bibitem[Ma et al. (1998)]{ma1998}
  Ma, C. et al. \ 1998, AJ, 116, 516
\bibitem[Menten \& Reid (1995)]{menten1995}
  Menten, K. M., Reid, M. J. \ 1995, ApJ, 445, 157
\bibitem[Perryman et al. (1995)]{perryman1995}
  Perryman, M. A. C., et al. \ 1995, A\&A, 304, 69
\bibitem[Perryman et al. (1997)]{perryman1997}
  Perryman, M. A. C., et al. \ 1997, A\&A, 323, L49
\bibitem[Petrov et al. (2007)]{petrov2007} Petrov, L., Horota, T., Honma, M., 
  Shibata, K. M., Jike, T., Kobayashi, H., \ 2007, AJ, 133, 2487
\bibitem[Rodr\'\i guez et al. (2005)]{rodriguez2005}
   Rodr\'\i guez, L. F., Poveda, A., Lizano, S., Allen, C. \ 2005, ApJ, 627, L65
\bibitem[Sato et al. (2007)]{sato2007} Sato, M. et al. \ 2007, PASJ, in press
\bibitem[Ulich \& Haas (1976)]{ulich1976}
   Ulich, B. L., Haas, R. W. \ 1976, ApJS, 30, 247
\bibitem[Xu et al. (2006)]{xu2006}
  Xu, Y., Reid, M. J., Zheng, X. W., Menten, K. M. \ 2006, Science, 311, 54
\end{thebibliography}
\end{document}